\documentstyle[preprint,aps]{revtex}

\begin{document}
\draft
\preprint{HUP-96-0321}
\title{ The deformed uncertainty relation and the \\
           corresponding beam quality factor\footnote{The project
supported by the National Natural Science Foundation of China and
Zhejiang Provincial Natural Science Foundation of China.}}

\author{Kang Li, Dao-Mu Zhao, and Shao-Min Wang }

\address{Department of Physics, Hangzhou University\\
Hangzhou,310028, P.R. China}
\date{Received  23 March, 1996}
\maketitle
\begin{abstract}
By using the theory of deformed quantum mechanics, we study the deformed
light beam theoretically. The deformed beam quality factor $M_q^2$ is given 
explicitly under the case of deformed light in coherent state. When the 
deformation parameter $q$ being a root of unity, the beam quality factor
$M_q^2 \leq 1$.
\end{abstract}

%\narrowtext
%\twocolumn

\section{ Introduction}

The beam quality factor $M^2$ is a very important concept in laser physics,
which expresses directly the goodness of a laser beam. So the theoretical
analysis and experimental measurement of the beam quality factor is a very
interesting and hot subject recently.

It is known that the beam quality factor can be defined as$^{[1]}$:
$$
M^{2}=\frac{\rm Real~ beam~ space- beam ~width ~product}
{\rm Ideal~Gaussian ~space- beam~width ~product}.
$$
Its mathematical expression is
\begin{equation}
M^2 =\frac{\pi}{\lambda}\theta .\omega_0 ,
\end{equation}
where $\theta$ is the far-field angular spread of the beam, $\omega_0$
is the beam waist radius. The theoretical value of $M^2$ can be calculated
by second moment method as:
\begin{equation}
\theta =2\pi\frac{\int\int (s_x-\bar{s}_x)\hat{I}(s_x ,s_y)ds_x ds_y}
{\int\int \hat{I}(s_x,s_y)ds_x ds_y}
\end{equation}
$$
\omega_0 =2\frac{\int\int (x-\bar{x})I(x ,y,z)dx dy}
{\int\int I(x,y,z)dx dy}\mid_{z=z_0}
$$
where $I(x,y,z)$ is a time-average intensity profile and $\hat{I} (s_x, s_y)$
is a spatial frequency distribution, $z_0$ is the beam waist location.
In the far-field condition, the spatial frequency and the divergence $\theta$
have the relation $s=\frac{\theta}{\lambda}$. Using the uncertainty relation
of quantum mechanics, the authors in reference [2] proved that for any beam
its beam quality factor defined by the second moment method can not be less 
than $1$.

On the other hand, if a beam propagates through a self-focusing nonlinear
medium, its effective divergence angle is$^{[3] [4]}$
\begin{equation}
\theta_{eff}^2 =\theta^2 -\beta J,
\end{equation}
where $\theta$ is the divergence angle for same beam propagating through
a linear medium, $\beta$ is a constant depending on the medium,
$J$ is a quantity related to the critical power of the beam.
From equation (1) we have:
\begin{equation}
M_{eff}^2 
=(M^4 -\frac{\pi^2}{\lambda^2}\omega_0^2 \beta J)^{\frac{1}{2}}.
\end{equation}
Obviously when $\beta J$ positive, $M_{eff}^2 <M^2$, namely when a 
beam propagates through a self-focusing nonlinear medium, the beam quality
factor will reduce. If the beam is the ideal Gaussian beam, then in the
nonlinear medium, its quality factor will less than $1$.

Recently, a new $CO_2$ laser with equivalent beam quality factor 
$M_e^2=0.3$ is realized
experimentally by the Hangzhou University and the 12th Institute of
National Electronic and Industry Department of China$^{[5]}$. 

According to the reference [2], these results of $M^2\leq 1$ can not be 
analyzed by the theory of quantum mechanics.

We call the beam being the deformed beam if the beam quality factor is less
than $1$. In this paper we will use the theory of q-deformed quantum
mechanics$^{[7,8]}$ to disciple the deformed beam, and give out the 
generalized
beam quality factor $M_q^2$ explicitly , where the deformation parameter
$q$ plays a row of a bridge connecting the usual beam and deformed beam.
The paper is organized as following: In second part, we give out the
deformed uncertainty relation in the q-deformed quantum mechanics.
In third part, we discuss the uncertainty and the deformed beam
quality factor in the deformed coherent state.
The deformed wave functions of deformed coherent states are given in
the part four. There are some results and discussions given in the last
part.

\section{ THE DEFORMED UNCERTAINTY RELATION IN q-DEFORMED
QUANTUM MECHANICS}

In recent years, on the base of quantum group there developed a new quantum 
mechanics theory--
q-deformed quantum mechanics$^{[8]}$. In the deformed quantum mechanics,
the derivative is replaced by a parametric derivative, and the Hamiltonian
has quantum group symmetry. In the one dimensional case, the coordinate and 
momentum of quantum mechanics (QM) and that of deformed quantum mechanics 
(DQM) have the following relation ship

\begin{center}
\begin{tabular}{|l|l|l|l|}
\hline   ~ & QM & DQM & Relation \\
\hline
{\rm Coordinate} & x & $x_q$ &$x_q=x$ \\
\hline
{\rm Momentum} & $-i\hbar\frac{d}{dx}$ &$-i\hbar\frac{d_q}{d_qx}$&$\frac{d_q}{d_qx}
=\frac{1}{x}[x\frac{d}{dx}]_q $\\
\hline
\end{tabular}
\end{center}

From the table above, we can see that in the q-deformed quantum mechanics
the coordinate does not deformed, but the momentum is deformed into
\begin{equation}
p_q =-i\hbar\frac{d_q}{d_q x} =-i\hbar\frac{1}{x}[x\frac{d}{dx}]_q
\end{equation}
where the q-bracket is defined by
\begin{equation}
[A]_q =\frac{q^A-q^{-A}}{q-q^{-1}}.
\end{equation}
It can be proven easily that the $[A]_{q=1}=A$, thus in this case $(q=1)$
the deformed quantum mechanics comes back to quantum mechanics.

Introduce operators:
\begin{equation}
a_q=\frac{1}{\sqrt{2\hbar}}(x_q+\hbar\frac{d_q}{d_q x})
\end{equation}
$$
a_q^{+}=\frac{1}{\sqrt{2\hbar}}(x_q -\hbar\frac{d_q}{d_q x}),
$$
then
\begin{equation}
x_q=\sqrt{\frac{\hbar}{2}}(a_q+a_q^{+}),~~~~p_q=-i
\sqrt{\frac{\hbar}{2}}(a_q -a_q^{+}).
\end{equation}
The operator $a_q,~a_q^{+}$ as well as another operator $N_q$
form a deformed harmonic oscillator algebra:
\begin{equation}
[a_q,a_q^{+}]_q =a_q a_q^{+}-qa_q^{+} a_q=q^{-N_q},
[N_q,a_q^{+}]=a_q^{+}, [N_q, a_q]=-a_q.
\end{equation}
$N_q$ and $a_q, a_q^{+}$ have the following relations:
\begin{equation}
a_q\cdot a_q^{+}=[N_q+1]_q,
a_q^{+}\cdot a_q =[N_q].
\end{equation}
From the equations (8),(9),(10), we obtain the commutator between
$x_q$ and $p_q$ as
\begin{equation}
[x_q, p_q]=i\hbar\{ [N_q+1]_q -[N_q]_q\}.
\end{equation}
If we define
\begin{equation}
(\Delta x_q)^2 =\overline{(x_q -\bar{x}_q)^2}=\overline{x_q^2}-
\overline{x_q}^2,
(\Delta p_q)^2 =\overline{(p_q -\bar{p}_q)^2}=\overline{p_q^2}-
\overline{p_q}^2,
\end{equation}
where the bars over the operators represent the quantum mechanics average
values. Then from the general principle of uncertainty relation, we have
\begin{equation}
(\Delta x_q)^2\cdot (\Delta p_q)^2\geq
\frac{\hbar^2}{4}\overline{([N_q+1]_q -[N_q]_q)}^2
\end{equation}
Reference [8] pointed out that, for the consistent q-deformed quantum
mechanics, the deformation parameter can only take the value in the range:
(1) q being positive real number, (2) q being a root of unity. When q being
positive, the deformed uncertainty is greater than that of undeformed case;
When q being a root of unity, $\overline{([N_q+1]_q -[N_q]_q)}^2\leq 1$, the
corresponding uncertainty is weaker than that of the quantum mechanics in 
general. In order to describe the deformed beam, we are interested very much
in the case of q being a root of unity.

\section{ THE UNCERTAINTY AND THE BEAM QUALITY FACTOR
          IN THE DEFORMED COHERENT STATE}

In the sequels, we choose q being a root of unity, i.e., $q=\exp (\frac{i\pi}
{p+1})$, then the Hilbert space corresponding to the deformed harmonic 
oscillator algebra has finite dimensions--$(p+1)$ dimensions, its basis is 
\begin{equation}
|n>=\frac{(a_q^{+})^n}{\sqrt{[n]_q}}|0>, ~~~n=0,1,2,\cdots ,p
\end{equation}
We can obtain easily that:
\begin{equation}
\begin{array}{l}
a_q^{+}|n>=\sqrt{[n+1]_q}|n+1>\\
a_q|n>=\sqrt{[n]_q}|n-1>\\
a_q|0>=0, a_q^{+}|p>=0, N_q|n>=n|n>.
\end{array}
\end{equation}
Or $a_q, a_q^{+}, N_q$ can be represented as
\begin{equation}
\begin{array}{l}
a_q=\sum_{n=1}^{p}\sqrt{[n]_q}|n-1><n|\\
a_q^{+}=\sum_{n=1}^{p}\sqrt{[n]_q}|n><n-1|\\
N_q=\sum_{n=1}^{p}n|n><n|.
\end{array}
\end{equation}

The coherent state $|\alpha >$ is defined as the eigenstate of $a_q$, i.e:
\begin{equation}
a_q|\alpha >=\alpha |\alpha >
\end{equation}
which can be expanded by the basis of the Hilbert space:
\begin{equation}
|\alpha >=\sum_{n=0}^{p}c_n |n>.
\end{equation}
Applying  $a_q$ to both sides of (18), we can get the recurrence relation of
$c_n$, and then obtain $c_n$ as
\begin{equation}
c_n =\frac{\alpha^n }{\sqrt{[n]_q !}}c_0 ,
\end{equation}
where $c_0$ is the normalization factor which is given by
\begin{equation}
c_0 =(\sum_{n=0}^{p}\frac{|\alpha|^{2n}}{[n]_q !})^{-\frac{1}{2}} .
\end{equation}
Set
$$
|\alpha_m >=(\sum_{n=0}^{p}\frac{|\alpha_m |^{2n}}{[n]_q !})^{-\frac{1}{2}}
\sum_{n=0}^{p}\frac{\alpha_m^{n}}{\sqrt{[n]_q !}}|n> ,
$$
then from the knowledge of linear algebra, we know that there are
$(p+1)$ independent $|\alpha_m >$ which satisfy
$$
\sum_{m=0}^{p}|\alpha_m><\alpha_m | =1.
$$

The relation between space frequency $s$ and momentum $p$ is
\begin{equation}
s=\frac{p}{2\pi\hbar}.
\end{equation}
So $\Delta s=\frac{\Delta p_x}{2\pi\hbar}$ . Then from equations (1) and (2),
we get that
\begin{equation}
(\Delta x)^2\cdot (\Delta p_x)^2 =\frac{\hbar^2}{4}M^2.
\end{equation}
In the following, we calculate the $(\Delta x)^2$ and $(\Delta p_x)^2$
for the deformed coherent state. From equations (8) and (17), we have:
\begin{equation}
\overline{x_q}=<\alpha |x|\alpha >=\sqrt{\frac{\hbar}{2}}(\alpha^*+\alpha )  
\end{equation}
\begin{equation}
\overline{x_q^2}=\frac{\hbar}{2} \left( \alpha^{*2} +\alpha^2 +\alpha^*\alpha
+\frac{1}{2}(q+q^{-1})|\alpha|^2 +\sqrt{1+\frac{1}{4}(q-q^{-1})^2
|\alpha|^4}\right) 
\end{equation}

\begin{equation}
\overline{p_q}=<\alpha |p_x|\alpha >=-i\sqrt{\frac{\hbar}{2}}
(\alpha^-\alpha^* ) , 
\end{equation}

\begin{equation}
\overline{p_q^2}=-\frac{\hbar}{2} \left( (\alpha  -\alpha^{*})^{2} 
+\alpha^*\alpha -\frac{1}{2}(q+q^{-1})|\alpha|^2 -\sqrt{1+\frac{1}{4}
(q-q^{-1})^2|\alpha|^4}\right) .
\end{equation}
Using the equation (12), we get
\begin{equation}
(\Delta x_q)^2=\overline{x_q^2}-\overline{x_q}^2
=\frac{\hbar}{2} \left(\frac{1}{2}(q+q^{-1}-2)|\alpha|^2 
+\sqrt{1+\frac{1}{4}(q-q^{-1})^2|\alpha|^4}\right) .
\end{equation}
\begin{equation}
(\Delta p_q)^2=\overline{p_q^2}-\overline{p_q}^2
=\frac{\hbar}{2} \left(\frac{1}{2}(q+q^{-1}-2)|\alpha|^2 
+\sqrt{1+\frac{1}{4}(q-q^{-1})^2|\alpha|^4}\right) .
\end{equation}
So:
\begin{equation}
(\Delta x_q)^2\cdot (\Delta p_q)^2 =
\frac{\hbar^2}{4} \left(\frac{1}{2}(q+q^{-1}-2)|\alpha|^2 
+\sqrt{1+\frac{1}{4}(q-q^{-1})^2|\alpha|^4}\right)^2 .
\end{equation}
From the equations (22) and (29), we find that the beam quality factor $M^2$
of deformed coherent state reads:
\begin{equation}
M_q^2=\mid \frac{1}{2}(q+q^{-1}-2)|\alpha|^2 
+\sqrt{1+\frac{1}{4}(q-q^{-1})^2|\alpha|^4} \mid .
\end{equation}
Obviously, when $ q=1, M_q^2=1$. Namely, the coherent in quantum mechanics is
the ideal Gaussian beam. When $q$ being a root of unity, 
$q-q^{-1}=2i\sin\frac{\pi}{p+1}, q+q^{-1}=2\cos\frac{\pi}{p+1}$, then
\begin{equation}
M_q^2=|\sqrt{1+|\alpha|^4\sin^2\frac{\pi}{p+1}}
-(1-\cos\frac{\pi}{p+1})|\alpha|^2|.
\end{equation}

Now we discuss two cases as following:

1), for a given coherent state, i.e. the value of $\alpha$
is given, in order to keep the $M_q^2$ being real, the integer
$p$ must be constrained as:
\begin{equation}
p\geq\frac{\pi}{\arcsin\frac{1}{|\alpha|^2} -1}
\end{equation}
When $|\alpha | =1$, $p$ can be any positive integers, under this case we have
\begin{equation}
M_q^2 =|2\cos\frac{\pi}{p+1}-1|
=\left\{ \begin{array}{ll}
1 & p=1\\
2\cos\frac{\pi}{p+1}-1 & p\geq 2
\end{array}\right.
\end{equation}
The table below shows that $M_q^2\leq 1$.

\begin{center}
\begin{tabular}{|l|l|l|l|l|l|l|l|l|l|}
\hline
p&1&2&3&4&5&6&7&$\cdots$&$\infty$\\
\hline
$M^2$&1&0&0.414&0.618&0.732&0.802&0.848&$\cdots$&1\\
\hline
\end{tabular}
\end{center}

2), for a given integer $p$ (i.e. the deformation parameter
$q$ is given). For example, for $p=3$, namely $q=e^{\frac{i\pi}{4}}$,
equation (31) becomes
\begin{equation}
M_q^2=|\frac{2-\sqrt{2}}{2}|\alpha|^2-\sqrt{1-\frac{1}{2}|\alpha|^4}|,
\end{equation}
and the constraint for $|\alpha|$ is
\begin{equation}
|\alpha|^2\leq\sqrt{2}.
\end{equation}
Under the constraint of (35), the $M_q^2$ given by equation (35)
takes values in the region from $0$ to $ 0.414$ which is less than $1$.

\section{THE EXPRESSION OF DEFORMED WAVE FUNCTION IN COHERENT STATE}

From equations (18),(19) and (20), we know that, in the coordinate 
representation, the wave function of deformed coherent state is
\begin{equation}
\Psi (x) =<x|\alpha >=
(\sum_{n=0}^{p}\frac{|\alpha |^{2n}}{[n]_q !})^{-\frac{1}{2}}
(\sum_{n=0}^{p}\frac{\alpha^{n}}{\sqrt{[n]_q !}})<x|n> .
\end{equation}
Define
\begin{equation}
\psi_n =<x|n>,
\end{equation}
then from $a_q|0>=0$, we have
\begin{equation}
\frac{d_q}{d_qx}\psi_0 +\beta^2 x\psi_0=0,
\end{equation}
where $\beta^2=\frac{1}{\hbar}$.  The solution to equation (38)
reads:
\begin{equation}
\psi_0=C_0\sum_{n=0}^{\infty}(-1)^n\frac{\beta^{2n}}{[2n]_q !!} x^{2n}
       C_1\sum_{n=0}^{\infty}(-1)^n\frac{\beta^{2n}}{[2n+1]_q !!} x^{2n+1}
\end{equation}
where
$$[2n]_q !!=[2n]_q[2n-2]_q\cdots[4]_q [2]_q,$$
$$[2n+1]_q !!=[2n+1]_q[2n-1]_q\cdots[3]_q [1]_q,$$
$C_0$ and $C_1$ are constants to be chosen by physics or mathematical 
conditions. When $q=1 (p\rightarrow\infty ),~
\psi_0\sim e^{-\frac{1}{2}\beta^2 x^2}$, which is just the result in 
quantum mechanics. When $p$ finite, $[p+1]_q=0,~[n(p+1)]_q=0$.
So, when $p$ being an even number, the second term of $\psi_0$ divergences,
we can choose $C_1=0$ to overcome the divergence. when $p$ being an odd number,
the first term of $\psi_0$ divergences, so we can choose $C_0=0$ in this case.
Now we have:
\begin{equation}
\psi_0(x)=\left\{\begin{array}{ll}
C_0\sum_{n=0}^{\infty}(-1)^n\frac{\beta^{2n}}{[2n]_q !!} x^{2n}& p~{\rm 
being~an~even~number}\\
C_1\sum_{n=0}^{\infty}(-1)^n\frac{\beta^{2n}}{[2n+1]_q !!} x^{2n+1}&p~{
\rm being~an~odd~number}
\end{array}\right.
\end{equation}

From $|n>=\frac{1}{\sqrt{[n]_q}}a_q^{+}|n-1>$ we have
\begin{equation}
\psi_n=\frac{1}{\sqrt{[n]_q}}a_q^{+}\psi_{n-1}
\end{equation}
From this recurrence relation, we can obtain
\begin{equation}
\psi_n=(\frac{1}{2^n [n]_q!})^{\frac{1}{2}}(\beta x-\frac{1}{\beta}
\frac{d_q}{d_qx})^n \psi_{0}(x),~~n=0,1,\cdots, p.
\end{equation}

\section{CONCLUSION REMARKS AND DISCUSSIONS}

In this paper, we give out the beam quality factor in deformed coherent
state from the q-deformed uncertainty relation. When $q=1$, $M_q^2=1$,
this is the result of quantum mechanics. When $q=\exp (\frac{i\pi}{p+1})$,
($p$ is finite integer), $M_q^2\leq 1$. So the deformed coherent state
corresponding to a deformed beam. The ideal Gaussian beam propagating
in the self-focusing nonlinear medium is just a deformed beam. If the
deformation part of $M_q^2$ of deformed Gaussian beam comes only from
the divergence angle $\theta_q$, namely:
\begin{equation}
M_q^2=\frac{\pi}{\lambda}\omega_0\theta_q ,
\end{equation}
then from the equations (31) and (43), we have
\begin{equation}
\theta_q= \frac{\lambda}{\pi\omega_0}|\sqrt{1-|\alpha |^4\sin^2
\frac{\pi}{p+1}}-2|\alpha |^2 \sin^2\frac{\pi}{2(p+1)}|.
\end{equation}
From the equations (3) and (44), the $\beta J$ coefficient of the medium
reads
\begin{equation}
\beta J= \frac{\lambda^2}{\pi^2\omega_0^2}|\alpha |^4(\sin^2\frac{\pi}{p+1}
+\frac{4}{|\alpha |^2}\sin^2\frac{\pi}{p+1}\sqrt{1-|\alpha|^4
\sin^2\frac{\pi}{p+1}}-2\sin^2\frac{\pi}{p+1} )
\end{equation}
Different medium corresponds to different deformed divergence angle 
$\theta_q$, that is to say, different medium corresponds to different 
deformation parameter. For the self-focusing medium, $0\leq \beta J\leq 1$,
so when a beam propagates in the medium,  the beam quality factor will
decrease.

Using the q-deformed quantum mechanics to study the deformed beam
which provides an effective method to study the diffraction free beam
in general, and make us know better about the inherent characteristics of the
beam propagating theory.

{\bf Acknowledgement}

The authors would like to thank Q. Lin and J.M. Xu for useful discussions.

\end{document}